# A Firefly Algorithm based for Power Management in Wireless Sensor Networks (WSNs)

Hossein Pakdel[1] . Reza Fotohi[2]

**Abstract** In Wireless Sensor Networks (WSNs), designing a stable, low-power routing protocol is a major challenge because successive changes in links or breakdowns destabilize the network topology. Therefore, choosing the right route in this type of network due to resource constraints and their operating environment is one of the most important challenges in these networks. Therefore, the main purpose of these networks is to collect appropriate routing information about the environment around the network sensors while observing the energy consumption of the sensors. One of the important approaches to reduce energy consumption in sensor networks is the use of the clustering technique, but in most clustering methods, only the criterion of the amount of energy of the cluster or the distance of members to the cluster has been considered. Therefore, in this paper, a method is presented using the firefly algorithm and using the four criteria of residual energy, noise rate, number of hops, and distance. The proposed method called EM-FIREFLY is introduced which selects the best cluster head with high attractiveness and based on the fitness function and transfers the data packets through these cluster head to the sink. The proposed method is evaluated with NS-2 simulator and compared with the Algorithm-PSO and Optimal clustering methods. The evaluation results show the efficiency of the EM-FIREFLY method in Maximum relative load and Network Lifetime criteria compared to other methods discussed in this article.

**Keywords** Wireless Sensor Networks (WSNs) . Power Management . Network Lifetime . Firefly Algorithm

✉ Hossein Pakdel*
   Hos.pakdel.eng@iauctb.ac.ir

✉ Reza Fotohi
   R_fotohi@sbu.ac.ir; Fotohi.reza@gmail.com

1  Department of Computer Engineering, Central Tehran Branch, Islamic Azad University, Tehran, Iran
2  Faculty of Computer Science and Engineering, Shahid Beheshti University, G. C. Evin, Tehran, Iran

# 1 Introduction

Because of the recent progress in the area of Micro-Electro-Mechanical-Systems (MEMS), wireless telecommunications, smart sensors, and digital electronics, low-power, small, and inexpensive sensor nodes can be built that have the capability of wireless communication. These small sensor nodes consist three elements: information processing, sensor, and wireless information transmission [1-4]. Wireless sensor nodes are powered by battery, which in most cases is non-rechargeable. Once the energy of the nodes is finished, the sensor network will fail. The main purpose of wireless sensor networks is primarily to collect accurate data and then, to maximize network lifetime using power management. To achieve the second goal, that is, to extend the network lifetime, the power consumption of the sensors should be as low as possible. In wireless sensor networks, nodes near the base station or a single head node loses their energy much faster because they are constantly transmitting data to other nodes in the network [5, 6]. It causes non-homogeneous power consumption in the network. The nodes located in the base station's transmission range are critical nodes. When the lifetime of the nodes located in critical area is ended, other nodes lose their communication with the base station and would not be able to send data packets to this station. Thus, the whole network would be inaccessible and remaining energy of other nodes would be wasted. Hence, use of mobile sink in this paper is proposed.

Most of the optimization algorithms are inspired by the biological systems, such as the Firefly community algorithm. This algorithm searches for the optimal solution of problem by modeling a set of behavior of fireflies and allocating some values relevant to fitness of location of each Firefly as a model for the amount of Firefly pigments and updating the location of worms in successive iterations of the algorithm. In fact, the two main phases of the algorithm in each iteration are the pigment-refreshing phase and the movement phase. Fireflies move to other fireflies with more pigments in their neighborhood. In this way, the set tends to a better solution within successive iterations [7].

This paper proposes a method called EM-Firefly to improve power in wireless sensor networks. In the proposed method, the network environment is divided into hexagons and the firefly algorithm is used to select the appropriate cluster head. In the firefly algorithm, the light intensity is replaced by the residual energy of a sensor node, so attractiveness is directly related to energy and inversely related to distance. The attractiveness of each sensor node is compared to that of its neighbor, and the nodes with the highest attractiveness line up. Then, for these nodes, the fitness function is performed by the sink, so that the most optimal node is selected as the cluster head based on the fitness function and the remaining energy criteria, signal to noise rate, number of HoPs, and distance. The selected cluster head is responsible for collecting data at each round from the cluster to the sink.

This paper is organized as follows: in Section 2, preliminaries is demonstrated. The related works on improvement of wireless sensor networks are reviewed in Section 3. Details of the proposed method are described in Section 4. The results obtained from the simulation are evaluated in Section 5, and conclusion is presented in Section 6.

# 2 Preliminaries

This section provides a detailed description of the Firefly algorithm subsections in detail.

## A. Firefly Algorithm

The firefly algorithm was introduced in 2008 based on the behavior and flashing patterns of fireflies in nature. This algorithm is a kind of metaheuristic algorithm, inspired by nature and randomness, which is used in almost all (SI) group intelligence methods. This algorithm belongs to the group of random algorithms, which means that NP-Hard fields of optimization, engineering as well as problems of a kind of random search are used to achieve a set of solutions. [8].

### 2.1. Luciferin Update

The amount of Luciferin in each warm in each iteration is determined by the location of the worm. Thus, the worm's current Luciferin amount is added in each iteration considering fitness value proportionally. In addition, in order to model the gradual decline over time, some of the current Luciferin is reduced by a factor of less than 1. Thus, the Luciferin update is given as Eq. (1):

$$\eta_i(m) = (1-S) * \eta_i(m-1) + \gamma J(X_i(m)) \tag{1}$$

Where $\eta_i(m), \eta_i(m-1), D(K_i(m))$, denote new amount of Luciferin, previous amount of Luciferin, and location fitness of worm $i$ in iteration $m$ of algorithm $S$ and $\gamma$ are constants for modeling gradual decline and fitness effect on Luciferin.

### 2.2. Movement of Fireflies

In movement phase, each worm moves as probabilities toward one of the neighbors that has higher Luciferin. Hence, the worms move toward the neighbors with higher brightness. Figure 2 indicates neighbors, concepts of decision radius, and sensing radius of worms. In the b section of this figure, worms are ranked based on amount of Luciferin. That is, the worm with number 1 has highest brightness. Decision radius $rd$ actually specifies the range in which the worms are neighbors, and they are compared with the respective worm for Luciferin level. Sensing radius $rs$ specifies the upper bound for decision radius. In fact, during the iterations of the GSO algorithm, the decision radius of the worms varies depending on their circumstances. However, in any case, the decision radius of each Firefly does not exceed its sensing radius. The sensing radius of the modeler is the maximum ability of fireflies to observe other fireflies. In Fig. 2, the four fireflies $a, b, c, d$ have a higher amount of Luciferin than the Firefly e, but the Firefly $e$ is only within the view range of the fireflies $c$ and $d$, and so there are two possible directions for selection to move towards a more glowing Firefly [8].

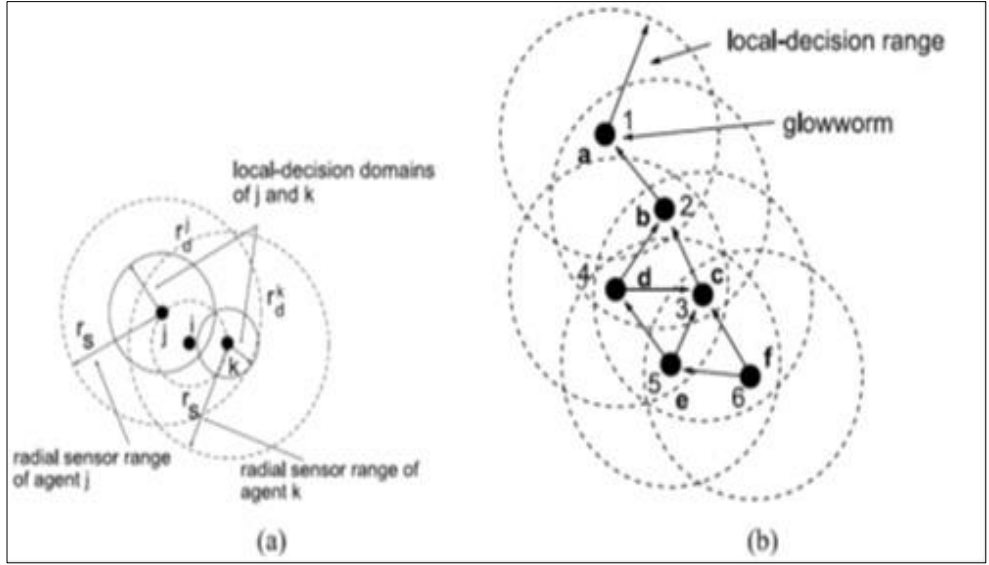

**Fig. 1**. A representation of the neighbors, the concepts of decision radius and the sensing radius of fireflies

Figure 1. (A) Representation of concepts of decision radius $rd$ and the sensing radius $rd$ For $k^{th}$ Firefly and $j^{th}$ Firefly toward $i^{th}$ Firefly: $(d(k,i) = d(j,i < rs, rdk >))$ therefore, since the Firefly i is within the sensing radius of both the k and j fireflies, but the decision radius of Firefly k is smaller than the required distance. So only Firefly j uses information of Firefly i. (b) Directional graph with respect to the ratio of Luciferin amounts of fireflies in the neighborhood, where the fireflies are arranged based on the amount of Luciferin and the brightest Firefly is represented by rank 1. For each Firefly i, the probability of moving to a brighter neighbor j is defined as Eq. (2):

$$S_{ij}(m) = \left( \frac{\eta_j(m) - \eta_i(m)}{\sum_{k \in N_i(m)} \eta_k(m) - \eta_i(m)} \right) \quad (2)$$

Where $N_i(m)$ denotes total fireflies in neighborhood of Firefly I at time m, $S_{ij}(m)$ is Euclidean distance between the Firefly i and j at time m, and $\eta_i(m)$ represents the variable neighborhood range related to the Firefly i at time m. Assuming selection of Firefly j by Firefly i (with the probability p that is obtained from Eq. (3), the motion-discrete time of Firefly can be written as follows:

$$K_i(m+1) = K_i(m) + S \left( \frac{K_j(m) - K_i(m)}{\|K_j(m) - K_i(m)\|} \right) \quad (3)$$

Where $K_i(t)$ is the m-dimensional vector of the location of the Firefly i at time t $\|K_j(m) - K_i(m)\|$ shows the Euclidean soft operator and s is the step size of the motion.

### 2.3. Neighborhood Update

A neighborhood is allocated to each Firefly $i$, radial range rid of which is naturally dynamic ( $0 < rid < rs$ ). The reason for not using a fixed neighborhood is justified in this way that since the fireflies just need local information of their neighborhood for movement, it is expected that the number of identifiable peaks be subject to the sensing radial range. In fact, if the sensing ranges of each Firefly covers the whole search space, all fireflies move toward the optimal global direction and local optimums are ignored. Since predefined information about the objective function (e.g., number of peaks or distance between peaks) is not assumed for the problem, it is simply not possible to consider a fixed neighborhood suitable for all objective functions. For example, the choice of the neighborhood range $rd$ is more suitable for the objective functions with the least distance between the peaks greater than $rd$ compared to the functions that this distance is less than $rd$ in them. The GSO therefore employs an adaptive neighborhood range for identifying the existence of multiple peaks in multi-modal function optimization problems. Assuming $r_0$ as the initial neighborhood range for each Firefly ($r_0 = rid(0)$), the neighborhood range of each Firefly is updated in each iteration of the GSO algorithm according to the following Eq. (4):

$$Y_d^i(m+1) = \min\left\{Y_s, \max\left\{0, Y_d^i(m) + \psi\left(n_t - |N_i(m)|\right)\right\}\right\} \tag{4}$$

Where $\psi$ is a constant, and $N_i(m)$ is a parameter for controlling the number of neighborhoods [9].

## 3 Related work

In this section, recent work on energy and power management in wireless sensor networks is described in detail. In most wireless sensor network applications, recharging the battery of the nodes is not possible. Thus, the protocols designed for these networks should be as energy-efficient as possible. Clustering and determining the transmission route to the base station in multi-step routing is one of the main approaches for designing scalable and energy-efficient protocols of wireless sensor networks. Clustering protocols are suitable methods for increasing network lifetime. However, most of these methods impose high power consumption on the head of each cluster so that protocol should change the arrangement of the clusters and head of each cluster in each period in order to increase the network lifetime. The other method is use of heuristic algorithms, such as Firefly algorithm or using ant colony algorithm.

In [10], to optimize energy consumption, a clustering approach is proposed for dynamic wireless sensor networks that ensures minimum energy consumption. This method is more powerful than traditional solutions because in traditional solutions the choice of cluster head was based on battery energy, but we went beyond this in this article. The proposed algorithm significantly affects the network reliability based on other parameters. Based on the results, it was found that the proposed algorithm can effectively and efficiently manage energy consumption between clusters. The main innovation of this method is that instead of direct routing to the base station, it uses UAV to collect data for routing, which increases the life time of sensor networks.

In [11], routing using Firefly algorithm in wireless sensor networks is discussed. All nodes declare their power consumption in the network, and each node compares the obtained energy with its own energy so that attraction level is obtained. Increasing periods of wireless sensor network lifetime using Firefly algorithm is described in [6]. This algorithm balances power of nodes and increases network lifetime. In this idea, routing is done using a spanning tree to transfer aggregated data. In all the works on Firefly algorithm, the sink is fixed and all the nodes in the network are aware of their location information, and this is a limitation for wireless sensor networks and questions its performance.

UAVs are used to collect data from locally distributed wireless sensors, which is a low-power solution. To this end, we minimize UAV energy consumption to gather the information needed during a tour. This tour uses both the energy from the movement of UAVs and the energy from data transmission. The purpose of this work is to determine the situations in which the UAV stops to collect information from each sensor cluster, as well as the route that the drone must take to complete its data collection tour. The proposed method faces two main challenges: the first challenge is to determine the stopping positions and the sensor subgroup is targeted to collect data from each subgroup using a clustering approach. In the second challenge, we use the remote vendor solution to find the path between these stops [12].

A solution to extend the network lifetime using controllable mobile heads and the LEACH algorithm is presented in [13]. Unintended and random movement of the cluster head nodes in the network will result in excessive overhead and energy loss. Therefore, the cluster head nodes should move in a controlled manner to the area where the data is actively sensed. In addition, in such algorithms as Low-Energy Adaptive Clustering Hierarchy (LEACH) and Link State Routing (LSR) for solving Uneven Energy Consumption (UEC) problem, different nodes are periodically selected as cluster head, which changing cluster heads causes imposing excessive overhead on system. In order to reduce the load of cluster head, two-cluster method was introduced. In these methods, two main cluster heads and auxiliary cluster head are used to collect data from the cluster members and send data from the lower cluster heads, respectively. The cluster head is selected based on the remaining energy, the ratio of the number of members in the neighborhood radius, and the strength of the received signal. The disadvantage of this method is to ignore the dimension of the distance and the constant switching of the cluster head. The selection criteria for cluster head are remaining energy and node density. The advantage of this method is the selection of auxiliary cluster head based on node density. In selection of the cluster head based on the remaining energy, the node distance to the main station and the frequency of algorithm run are considered. The disadvantage of this method is lack of balance in distance between members and cluster head.

In [14], fuzzy logic based on specific fuzzy descriptors is used to select the cluster head. Particle swarm optimization algorithm is used to optimize fuzzy membership functions to improve their amplitude. In this paper, the sensor network is divided into clusters and the moving collector starts from a fixed sink or base station and moves through each of these clusters, collecting data from the selected clusters in a single hop. he does. To prove the effectiveness of this method, it is inspired by the method based on ant optimization in which the cluster head is selected based on demand instead of clustering in each round.

In MLS, a hypothetical node is considered as the root and other nodes are added individually to the spanning tree. This algorithm is somehow similar to Dijkstra's algorithm, and it is the shortest path. The major difference between MLS and Dijkstra's algorithm is that in MLS, there is no need for the new node to have the shortest path to the root among existing nodes in each stage that the new node is added to the spanning tree, but it should avoid the nodes that currently have heavy loads. Running computations in the presented algorithm for selection of a node with lower load causes high energy consumption in network [15].

They proposed a new approach to wireless sensor networking for PCA-based data reduction based on a predictive model-based dataset. In this method, additional data is removed and energy consumption in sensor networks is reduced. To define the data prediction model in the CH node, the ARIMA model is used, which saves a significant amount of energy in the wireless sensor network. This ARIMA-based forecasting model is sent to other cluster nodes and the nodes predict the event based on the data model. Because each node transfers the deviations between the original data and the predicted data instead of transferring the original data to the cluster head node. No data is transmitted by the sensor node if the deviation is below a certain threshold. Therefore, the number of data transmitted by sensor nodes is significantly reduced [16].

Energy-Efficient PEGASIS-Based protocol (IEEPB) provides a new method for developing chain and is composed of three phases. First phase is the phase of chain development where the network parameters are valued. The number of network nodes, primary energy, situation of base station, etc. are determined. The base station sends a Hello Packet to the whole network to obtain initial information of network, including living nodes' ID and distance of nodes to base station. The nodes farther away from the base station are taken as the end node and tagged with node 1. Each node in the information chain acquires its distance from other nodes not yet connected to the chain, naming the closest node to itself with i value. The i value represents the ith node in the chain. The node i obtains its distance to the i-1 node (which is the node within the chain). Then, it identifies its nearest node and connects it to the chain ($1 \leq j \leq i-1$). In this time, node i is recognized as the last node of the new chain. This process continues until all nodes are connected to the chain. IEEPB determines the leader node considering remaining energy of node and node distance to the base station. The data transfer method is similar to the one used in public PEGASIS and data collection starts at the end of the chain, and each node gives its data to its nearest neighbor node so that the data is transmitted to the leader and finally to the base station. The IEEBP has more live nodes and more energy efficiency than public PEGASIS [17].

In [18], an Optimal Clustering method is proposed that models the optimal clustering problem as a separable convex optimization problem to obtain the optimal clustering size and optimal transfer radius. In this paper, we design a distributed clustering algorithm that builds the structure of inter-cluster data collection around virtual cluster heads in a wireless sensor network. The proposed algorithm is a cluster head linker based on the Hilbert-like curve to collect compressed sensed data between cluster heads in a common and cumulative manner. The simulation results show the superiority of the proposed method in the criteria of average reduction in energy consumption.

In general, PEGASIS, each chain has only one leader or cluster head. Considering limited energy of nodes, it is problematic when the cluster is large. In PDCH, two cluster head is considered for each chain so that formation of long chains is prevented and more energy balance happens in clusters. PDCH in each level just allows the nodes belonging to the main cluster to be cluster head. These cluster heads are defined as the main cluster head. Nodes belonging to the cluster head cannot select any cluster head as the second cluster head in the chain. In other words, it would differently behave with sub-clusters than the main clusters. If a level lacks sub-cluster, the same single cluster head used in general PEGASIS is used. The main cluster head is responsible for receiving data from other nodes and combining them and sending them the second cluster head. Second cluster head gives collected data of its level to the upper-level cluster head. Figure 3 indicates performance of this algorithm well. As observed in Figure 2, task of a cluster head in PDCH method is assigned to two cluster heads. This method allows the cluster heads to work longer and the energy efficiency will be significantly increased [19].

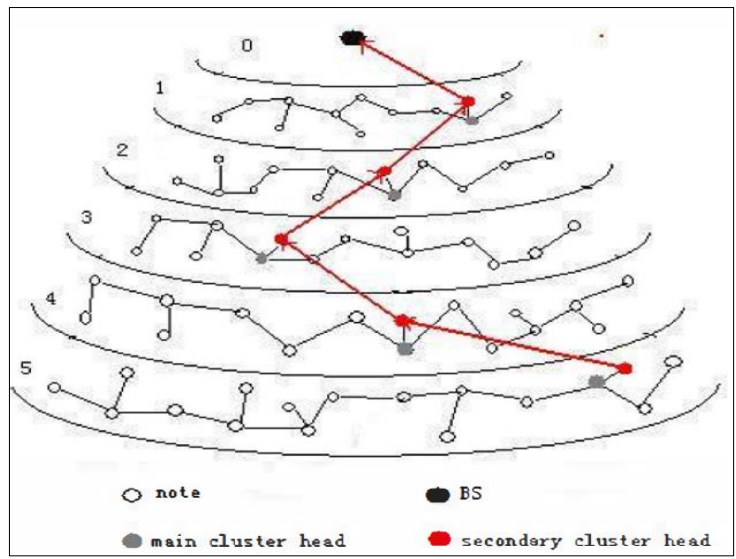

**Fig. 2**. Data aggregation in PDCH.

Zhao et al. studied the problem of increasing the lifetime of the network in large-scale wireless sensor networks. In this case, a mobile sink gathers information periodically over a pre-specified path, and each sensor node transfers its data to the mobile sink. They presented a heuristic topology control algorithm with time complexity O (n (m + n log n)) using the Grading program and dynamic programming, where m and n and m are the number of network edges and network nodes. The suggested approach is known as the Minimum Load Sets (MLS) algorithm. The suggested algorithm provides the tree-based heuristic control for maximization of the network lifetime in WSNs with the mobile sinks. The MLS specifies the parent node of each node through growing a tree from a root node. At any time, the MLS adds an edge to the growing tree for reducing the relative load of the nodes, which currently have the maximum axial load while minimizing other nodes [20].

## 4 Proposed approach

Because of specific characteristics of wireless sensor networks, there are various challenges in these networks. One of these challenges is limited energy source of nodes. Thus, presenting energy-efficient methods that increase lifetime of nodes and sensor networks has always been considered by the researchers. Clustering-based routing methods are among the main power consumption reduction methods in wireless sensor networks.

The cluster formation, the choice of cluster heads from the cluster's nodes, and the allocation of the tasks to it in these protocols considerably affect the network scalability, extending the network lifetime and energy efficiency. Moreover, in clustering-based approaches, the number of data transferred to the base station is decreased by the application of data synthesis techniques and by removing duplicate data. Additionally, the necessity of efficient optimization of network resources for extending the lifetime of large-scale, dense wireless sensor networks urged the researchers to investigate and present efficient clustering approaches. It has been confirmed that clustering is an effective way for organizing a large-scale WSN network into connected groups for increasing the reliability and lifetime of these networks. In the present work, a new algorithm based on Firefly is suggested. The suggested algorithm chooses a cluster head based on the position and energy level of the nodes related to the base station as a repository for the information gathered via Firefly algorithm. This algorithm presents stronger coverage by obtaining general knowledge at the base station that ensures accessibility of all nodes through the connected cluster head. In the suggested approach, the base station (BS) computes the number of times a cluster head can stay as a head. This feature largely influences extension of network lifetime since it can decrease the amount of energy wasted in the cluster head replacement.

### 4.1. Network assumptions in the proposed method

The sensor network consists of nodes that are randomly distributed. The following specifications are considered for our EM-Firefly method:

- The sensor nodes are homogeneous and energy-limited.
- The sinks have no energy limitation and their location in the network is fixed after the fixed network is established.
- The energy consumed for sensing and data processing is not considered.
- Sensors in the network are aware of their location
- Sensors always sense around them and send data to the cluster head at a fixed rate.
- The cluster head collects cluster information and gives to sink. Then, sinks transmit data to base station through communication or directly.
- Each period of data collection and transfer aggregated data from all cluster heads to sink is considered as one period.

### 4.2. Details of proposed EM-Firefly method

In clustering-based sensor network routing, each node provides the cluster head with some information. Then, it is transmitted to the sink from the cluster head, and then information is transferred to sink. In return, information is transferred in the same but in an inverse order (Figure 3).

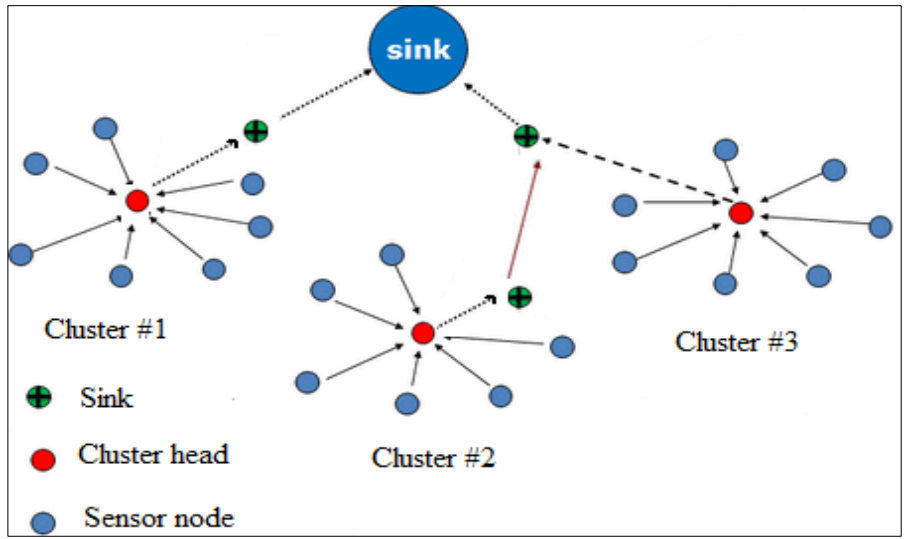

**Fig. 3**. Routing in clustering-based WSN.

The proposed method is called EM-FIREFLY that improves energy in sensor network using Firefly algorithm. In this method, firstly, the information of the whole network is collected, and using Firefly algorithm, the best cluster head is selected based on this information and energy criteria. The proposed EM-FIREFLY method contains three phases, which are described in the following.

### 4.2.1. First Phase: Network Clustering

In the proposed method, the first step is drawing a hypothetical circle with a radius of "distance of the farthest node to the central station" and thus all nodes will fall into our hypothetical circle. In the next step, the entire network area will be divided into several segments. In most cases, dividing the network area into six segments would be the most desirable segmentation mode. The number of segments can also be determined based on the number of optimal clusters in the system. This depends on various parameters such as network topology and the ratio of the computing cost to communications in the network. In the proposed method, we also use six sectors. In this model, the circle radius should be drawn in an angle that corresponds to the diameters of hexagon. Each segment has a sink and one cluster head must be chosen for each cluster. The operation of cluster head selection is performed by a sink.

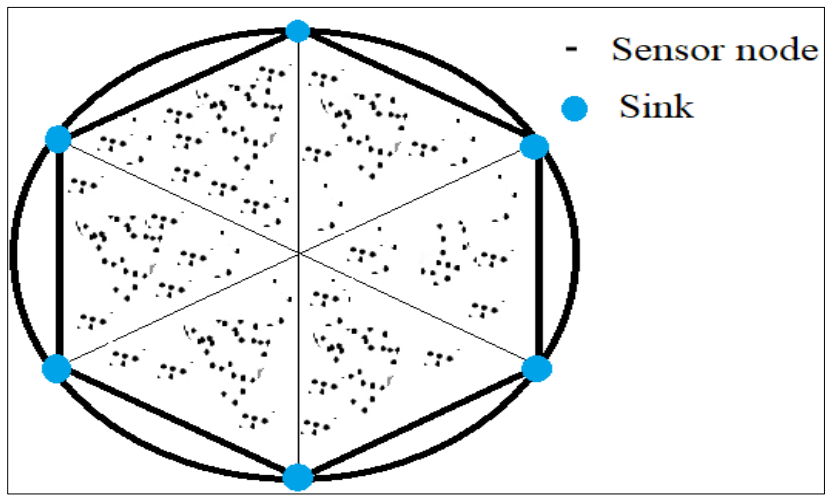

**Fig. 4**. Model of clustering in proposed EM-FIREFLY method.

### 4.2.2. Second Phase: Network Information Collection

In this phase, information of each cluster is collected in the proposed EM-FIREFLY method. For selection of the best cluster head out of the important routing parameters in the sensor networks, parameters of remaining energy, HoP count, distance, and signal-to-noise ratio (SINR) were considered. The higher the node's energy, the higher the probability of being alive and active for a longer time and the higher node's transmission range. The signal-to-noise ratio is introduced as SINR, and the larger it is, the higher probability of continuing bonding for longer time. The smaller HoP count and distance, the higher probability of sound bonding. To collect information, all nodes member of a cluster sends their information, including "geographic location" and "energy", in response to the HELLO message sent by the sink. Remaining energy parameter: The remaining energy in a node is a criterion that is directly related to its value; the more energy a node has, the wiser it is to select the node for the cluster head because packets cannot be transferred by the cluster head if the node runs out of energy. The remaining energy is calculated using the Eq. (5).

$$R_e(t) = P_e(t) - Cu_e(t) \tag{5}$$

In Eq. (5):

$R_e(t)$: Is remaining energy.

$P_e(t)$: Is primary energy.

$Cu_e(t)$: Is current energy.

**HoP count:** HoP count parameter is inversely related to the node value; the greater the HoP count, the greater the likelihood that the cluster head is inappropriate.

**Distance criterion:** The distance between the nodes in communication with each other at every certain time is obtained. Distance is a critical factor since the energy consumed in the packet transmission directly depends on the distance between nodes. The Eq. (6) is used for calculation of the distance.

$$\text{distance between two nodes}(r) = \sqrt{(X_1 - X_2)^2 + (Y_1 - Y_2)^2} \tag{6}$$

**Noise ratio (SINR):** Noise ratio is defined as the ratio of signal power (S) to a combination of noise (N) and interference (I) power, and it can be represented mathematically as Eq. (7).

$$SINR = \left(\frac{S}{N+I}\right) \tag{7}$$

As the sum of interference and noise power cannot be calculated exactly, SINR is estimated using the average receipt during the rest period. SINR is used to determine the quality of network connections, and the higher value indicates that the node has higher signal power. By sending a

HELLO message to all nodes, the sink maintains the information in its table based on the ID of each node. The information in this table includes identifying all the neighbors, the distance to the neighbor node, the amount of remaining energy of nodes, the SINR, and their HoP count.

### 4.2.3. Third Phase: Application of the Firefly Algorithm

Fireflies produce short-lived light through a process called bioluminescence. They use this method to attract prey or partner or to warn against a predator. Therefore, light intensity is an important parameter for Firefly insects. The Firefly algorithm follows three rules:
1) Firefly of any sex can attract another Firefly. In other words, all fireflies are bisexual, so that one Firefly can attract other fireflies, regardless of the sex.
2) An attraction factor is considered that depends on the brightness of the light as the fireflies move towards more attractive fireflies.
3) Brightness of Firefly is calculated through an objective function.

In the Firefly algorithm, light intensity and attraction are two important variables. The fireflies are absorbed into the Firefly that is brighter than it. Thus, the attraction depends on the light intensity.

**Algorithm 1:** Pseudo code for Firefly algorithm

1: ***Procedure*** Parameterization
2: **Let** *Max Generation*: Maximum number of generations
3: **Let** $\gamma$: Light attraction factor
4: **Let** $r$: Specific distance to light source
5: **Let** $d$: Range distance
6: Objective function $f1(y)$ where $y = (y_1,...,y_d)$
6: Production of early populations of fireflies or $y_i (i = 1, 2, 3, ..., m)$
7: **Determining** light intensity $I_i$ in $y_i$ through $f1(y_i)$
8: **While** (t <Max Generation)
9:     **For** i = 1 to m (For all $m$ fireflies)
10:     **For** j=1 to m
11:       **If** ($I_j > I_i$)
12:         Firefly i moves toward Firefly j
13:       **End if**
14:     Attraction with distance r changes through $EXP[-\gamma r^2]$.
15:     New evaluated solutions and light intensity are updated.
16:    **End for**;
17:  **End for**;
18:  Fireflies are ranked and the best current Firefly is found.
19: **End while**.
20:  Results are presented.
22: ***End Procedure***

The intensity of light defined as attraction is inversely proportional to the distance r from the light source. This shows that the attraction increases with decreasing distance and vice versa. The light intensity is calculated according to Eq. (8) [10].

$I$ : Intensity of light

$I_0$ : Intensity of primary light

$\gamma$ : Light attraction factor

$r$ : The distance between two fireflies

Light intensity id related to attraction and it is represented by β through Eq. (8) [11]:

$$e^{-\gamma r^2} \beta = \beta_0 \tag{8}$$

$\beta_0$ : Attraction in $r = 0$

The Cartesian distance formula is used for calculating distance between two fireflies:

$$r_{ij} = \|x_i - x_j\| = \sqrt{\sum_{k=1}^{d}(x_i,k - x_j,k)^2} \tag{9}$$

Where $k$, $x_i$, is the kth element of $x_i$ spatial coordinates related to $i^{th}$ Firefly. In the two-dimensional case, the same Eq. (10) is used to calculate the distance. The movement of the less bright Firefly $i$ to the brighter Firefly j is as Eq. (10):

$$\Delta x_i = \left(\beta_0 e^{-\gamma r_{ij}^2}(x_j - x_i) + \alpha\left(rand - \frac{1}{2}\right)\right) \tag{10}$$

In equation (4-7), if the light attraction factor(γ) approaches zero, then the attraction is almost constant, otherwise, if the factor approaches infinity, the attraction decreases. In this formula, the first term is for attraction, while the second term, α, is the random parameter. rand is a random number generator, which is uniformly distributed over the interval [0,1]. For most of our implementation, $\beta_0 = 1$ and $\alpha \in [0,1]$ can considered.

In the proposed EM-FIREFLY method, for selecting the appropriate cluster head in the algorithm described above, the light intensity is replaced by the remaining energy of a node. Therefore, the attraction is directly related to the energy. $x_j$ and $x_i$ is the distance between two nodes $x_j$ and $x_i$ that is inversely related to the light intensity.

Hence, firstly, the nodes are clustered in the network according to the first phase, and each node in the cluster sends the information related to its remaining energy as well as its position to the sink in response to the HELLO message from the sink. The sink firstly calculates remaining energy, distance of nodes in cluster, HoP count, and SINR based on the received message. Then, it calculates the attraction for the cluster nodes based on the Firefly algorithm on the basis of remaining energy and distance of each node. At this stage, an ascending queue is considered as the number of nodes. The attraction of each node is compared to the previous node in the queue, and if its attraction is higher, the node enters the queue, and the node with less attraction is eliminated from the queue. In the proposed method, the queue is considered for the three nodes, and three nodes with highest attraction remain in the queue to compute the fitness function for them. Therefore, the objective (fitness) function for nodes with the highest attraction is calculated according to Eq. (15). Finally, the nodes with highest fitness function are selection as the cluster head for sending cluster information and collecting data to be sent to the sink. The variable normalization should be done for calculation of fitness function so that the function can be calculated. For this purpose:

$$E = \left( \frac{R_{e,i}}{\sum_{i=1}^{N} R_e} \right) \quad (11)$$

; remaining energy of node i to sum of remaining energy of all nodes in the cluster.

$$S = \left( \frac{SINR_i}{\sum_{i=1}^{N} SINR_i} \right) \quad (12)$$

; The signal-to-noise ratio of node i to the sum of the SINR of all nodes in the cluster.

$$R = \left( \frac{r_{i,j}}{\sum_{i=1}^{N} r_{i,j}} \right) \quad (13)$$

; distance between two nodes to the sum of the distances between all nodes in the cluster.

$$H = \left( \frac{HOPcount_{i,j}}{\sum_{i=1}^{N} HOPcount_j} \right) \quad (14)$$

; HoP count between two nodes to the sum of HoP counts between all nodes in the cluster..

$$f_i = \sum_{i=1}^{N} \left( (W_1 * E) + (W_2 * S) + (W_3 * (1-R)) \right) + (W_4 * (1-H)); \quad (15)$$
$$W_1 + W_2 + W_3 + W_4 = 1$$

The sum of the coefficients $W_1$ to $W_4$ should be equal to one. Then, the factor of an agent such as energy can be considered above the other criteria. The value of these factors is determined in the simulation.

The selected cluster head is introduced to the entire cluster members via the sink. All nodes that are member of the cluster provide the cluster head with their collected data. So that the data is delivered to the sink through the cluster, head and then transferred to the base station or sink.

After each round of data collection, this information is updated in each node and the selection of the cluster head is performed again. In the Firefly-based method, remaining energy plays a significant role because its value is shared by other nodes in the network. The nodes with less energy are attracted by the node with higher energy, and an attraction factor is calculated. HoP count, SINR, and distance between two nodes in cluster are calculated. Based on these values and remaining energy, a new cluster head is found. This cluster head is the best in terms of energy, distance, etc. and causes that nodes pass smaller step and distance to transmit data. Thus, less energy is consumed, leading to energy and power management in the network and increasing network lifetime. The flowchart of the suggested steps is shown in Figure 5.

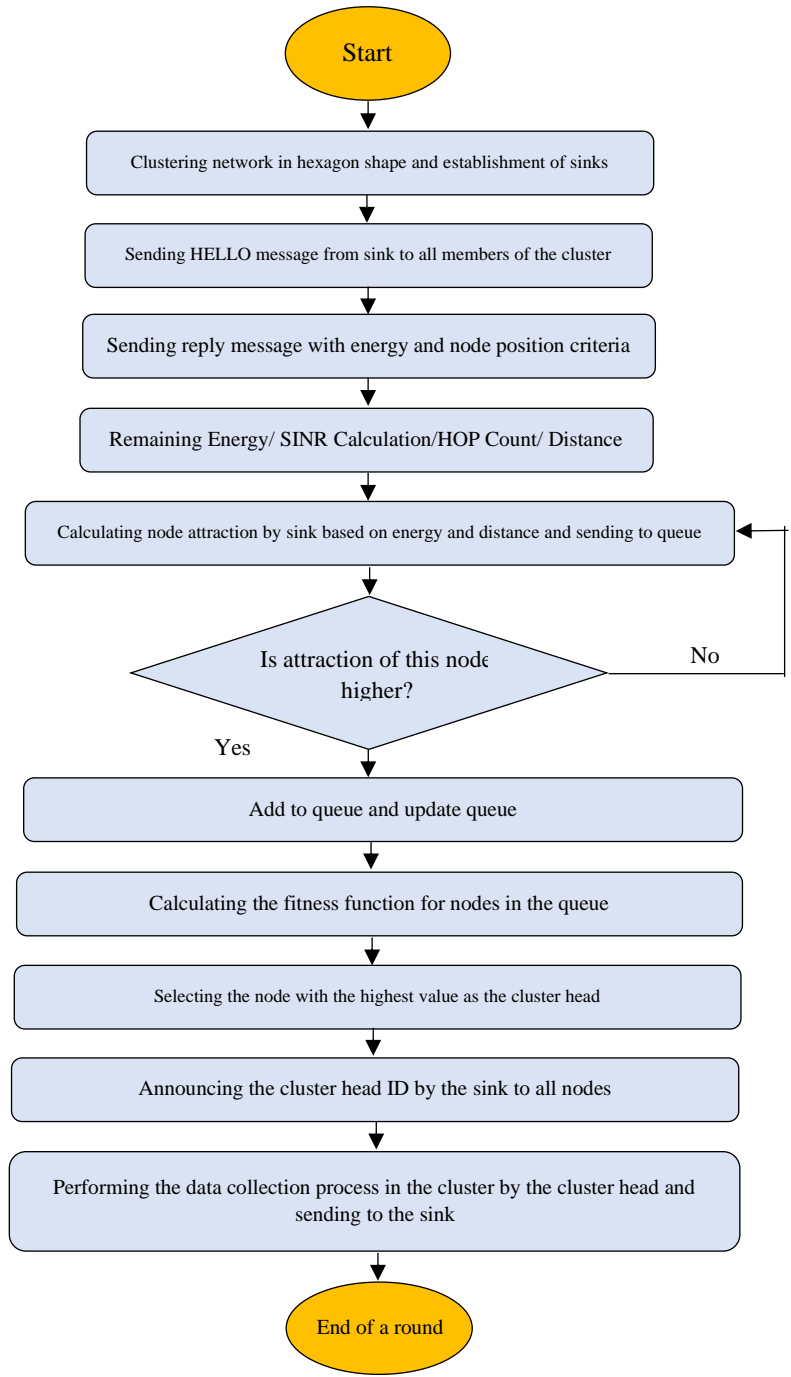

**Fig. 5**. Flowchart of proposed EM-FIREFLY method.

# 5 Experimental results

In this section, the proposed EM-Firefly method is evaluated using NS-2 stimulator, and the most important parameters of routing, such as maximum relative load and network lifetime, are used for measurement of efficiency of the proposed method. In order to indicate capability of the proposed method, it is compared with the optimal clustering method [10] and Algorithm-PSO [14] approaches described in Section 2.

### 5.1. Evaluation Criteria

**Maximum relative load:** Maximum relative load of node in sensor networks means ratio of its transfer to remaining energy [21].

**Network lifetime:** It is the duration when the network is stable and can sends data packets [22-25].

### 5.2. Simulation Environment and Scenario Design

In this paper, considering required research works and review of various papers on simulation software, it was concluded that the best simulator for wireless sensor networks in terms of flexibility and efficiency of software is NS-2 simulator, which is regarded as highly powerful simulator for sensor networks. This simulator is used for simulation of C++ codes. Various types of methods can be implemented and simulated and network performance can be observed using this type of simulator. We also used NS-2 simulator for simulation. Table 1 gives the values of the parameters used in simulation [26-28].

**Table 1** Simulation parameters

| Parameter | Value |
|---|---|
| Simulator | NS-2.35 |
| Number of sensor nodes | 4000 |
| Transmission range | 250 m |
| Antenna type | Omenia Antenna |
| MAC layer | 11_802 |
| Traffic type | CBR (UDP) |
| Buffer size | 150 packets |
| Positioning of nodes | Random |
| Methods | Optimal clustering method, Algorithm-PSO, and EM-Firefly |

## 5.3. Description of Simulation Scenarios

Table 1 gives the parameters required for simulation of the proposed method. Some of parameters, such as sensor node number, can be changed in each simulation run results obtained from simulation can be evaluated. The number of nodes available in the network was considered as 500 to 4000 nodes in several scenarios. The simulation environment used is the radio emission range as 250 for each node, and its MAC layer protocol is IEEE 802.11. There are also two traffic flows in the network that send the packet to the network at a fixed rate. We performed the simulation in three different scenarios. In the first scenario, only the number of the network node is assumed as variable, the hypothetical circle radius is equal to 20, and the energy of the nodes is considered as 200. The second scenario is based on the network model, which, as stated, is segmented and clustered in the proposed method based on the distance of the farthest node to the central station as a circle and then a hexagon. In the scenario, the circle radius varies based on the distance of the farthest node to the sink, which causes small or large size of circle. The higher is number of nodes, the radius of the hypothetical circle is considered as larger. Hence, in this scenario, by varying the number of nodes, the radius of the circle is varied from 5 m to 40 m. In the third scenario, the primary energy of the sensor nodes is variable, ranging from 50 to 400 J. We once considered these scenarios on our proposed method (EM-Firefly) and compared them with the proposed MLS method. The buffer size used in these scenarios is 150 packets. The position of the nodes in this scenario is considered as random.

## 5.4. Simulation results

Simulation was conducted on the proposed method, and it was ensured that the proposed EM-Firefly method works properly. Simulation results indicated that EM-Firefly outperforms in terms of maximum relative load and network lifetime criteria.

**Maximum relative load:** In wireless sensor networks, the relative load of the node means the ratio of its transmission cost to the remaining energy. Fig. 6 (a-d) shows the relative load comparison of the proposed method with Algorithm-PSO and Optimal clustering methods in all three scenarios. The maximum relative load for EM-FIREFLY is significantly lower in all three scenarios compared to the two methods compared. The reason for this is that the proposed method uses clustering to send data, and the selection of the cluster based on fireflies is such that the most important criterion is energy and distance from neighbors. When the best node is selected as the header in each round, it causes the load transfer between the nodes to be distributed in each round and also increases the amount of energy remaining in the nodes. Therefore, the maximum relative load is reduced. When the SINR is high and the number of steps and the distance from the head to the nodes is low, less energy is required for transmission, which in turn increases the residual energy in the nodes and reduces the relative load of the nodes in the network. The Algorithm-PSO method uses the PSO algorithm, although this reduces the relative load, but compared to the proposed method, this improvement is less, which is shown in the following figures that our proposed method, EM-FIREFLY performance It is better with different number of sensor nodes in the second scenario because as the number of sensor nodes increases, the radius of the circle increases compared to the first scenario and therefore the distance increases. It is slightly more overhead than the first and third scenarios, although still compared to Algorithm-PSO and Optimal clustering methods are improved. In the third scenario, the overhead is much better than the other two scenarios by increasing the energy of the nodes.

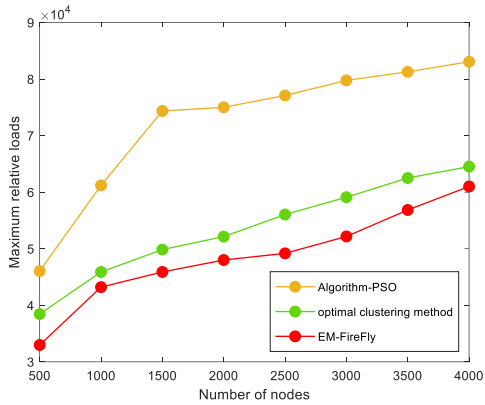

(a)  Maximum relative loads in the first scenario

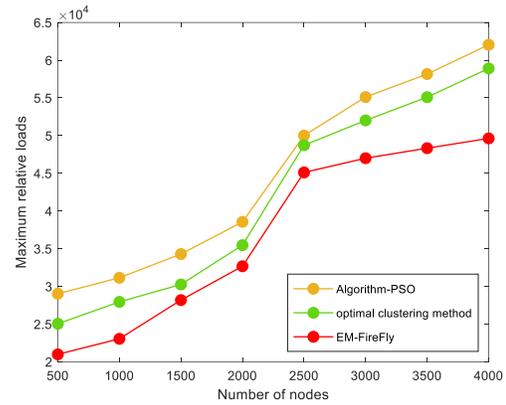

(a)  Maximum relative loads in the second scenario

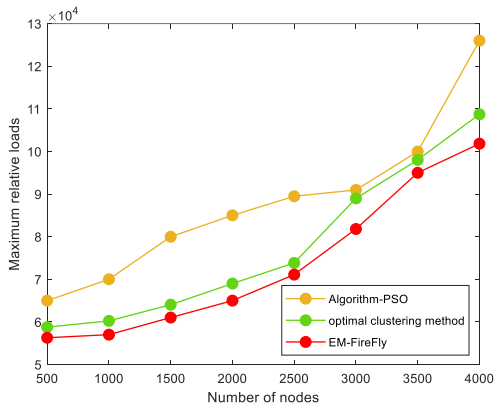

(c) Maximum relative loads in the third scenario

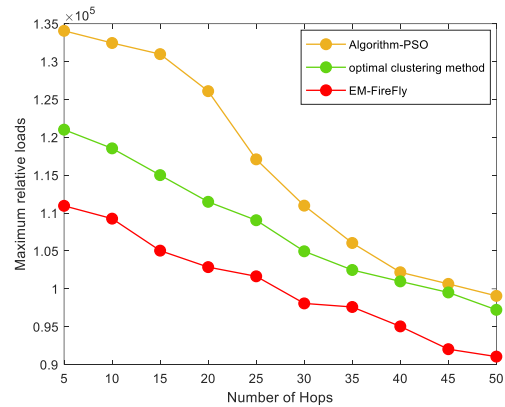

(d)  Maximum relative loads in the fourth scenario

**Fig. 6**. Maximum relative load in four scenarios.

**Network Lifetime:** Network lifetime is inversely related to the amount of energy consumed in wireless sensor networks. That is, with increasing energy consumption, the life of the network decreases. Therefore, power and energy management in the network depends on the amount of energy remaining in the sensor nodes, which directly affects the life of the network, and the better the power and energy management in the network, the amount of energy remaining in the node. Network capacity increases and network life increases. The simulation for this criterion is performed by changing the simulation time from 10 to 180 seconds. As you can see in Figure 7, the proposed EM-FIREFLY method also performs better than the Algorithm-PSO and Optimal clustering methods in terms of network longevity. In EM-FIREFLY algorithm, in each round, the nodes with the highest energy and attractiveness are first selected and then the best node is selected as the header, the best in terms of energy criteria, signal-to-noise rate, number of steps and distance. Considering these criteria, the most suitable node that has a good distance from its neighbors is selected as the header and less energy is used to transfer data. Therefore, the network life compared to Optimal clustering and Algorithm-PSO methods is increased by about 25 and 35%.

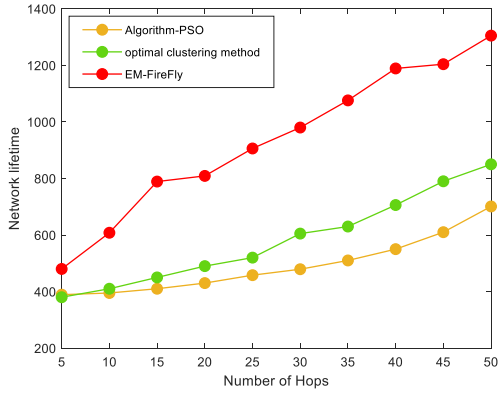

(a) Network lifetime in the first scenario

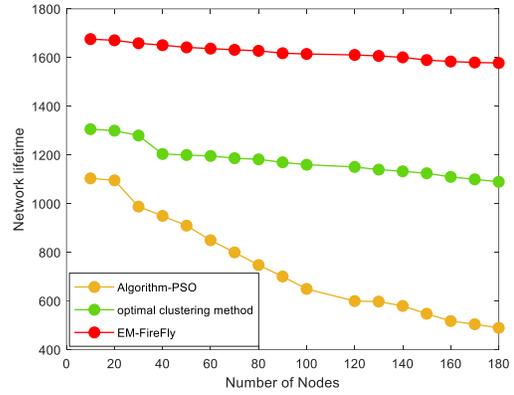

(b) Network lifetime in the second scenario

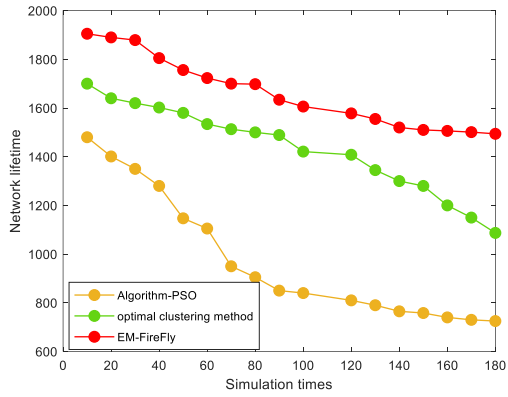

(a) Network lifetime in the third scenario

**Fig. 7**. Lifetime in four scenarios.

# 6  Conclusion

In this paper, we present an approach called EM-FIREFLY to collect common data based on optimal clustering to improve power in wireless sensor networks. In the proposed method, clustering and division of the network environment into hexagons have been used. To select the appropriate cluster head in the firefly algorithm, the light intensity is replaced by the residual energy of a node, so attractiveness is directly related to energy and inversely related to distance. The charm of each node is compared with its neighbor, and the nodes with the highest charm enter the queue. For these nodes, the fit function is performed by Sink, so that the most optimal node is selected as the heading based on the fit function and the remaining energy criteria, signal rate to noise, number of hops, and distance. The selected cluster head is responsible for collecting data in each round from the cluster to the Sink. The NS-2 simulator was used to extract the data and compare the proposed EM-FIREFLY method with the Algorithm-PSO and Optimal clustering methods. The evaluation results show that the proposed method performed better in the criteria of Maximum relative load and Network Lifetime than the other methods discussed in this paper.